\documentclass[11pt]{article}

\usepackage[left=2cm,right=2cm,top=2cm,bottom=2cm]{geometry}
\usepackage[utf8]{inputenc}
\usepackage[english,french]{babel}
\usepackage[T1]{fontenc}
\usepackage{amsmath}
\usepackage{amsfonts}
\usepackage{amssymb}
\usepackage{makeidx}
\usepackage{graphicx}
\usepackage{color}
\usepackage{enumerate}
\usepackage{enumitem}
\usepackage{multicol}
\usepackage{amsthm}
\usepackage[all]{xy}
\usepackage{bbm}
\usepackage{dsfont}
\usepackage[round]{natbib}
\usepackage{subfigure}
\date{}

\newcommand{\R}{\mathbb{R}}
\newcommand{\N}{\mathbb{N}}

\newcommand{\indi}{\mathbbm{1}}

\renewcommand{\phi}{\varphi}

\newtheoremstyle{complement}
   {4ex}
   {4ex}
   {\scriptsize}
   {5ex}
   {\itshape}
   {.}
   { }
   {}

\theoremstyle{definition}
\newtheorem{defi}{Definition}[section]

\theoremstyle{plain}
\newtheorem{theo}{Theorem}[section]

\newtheorem{pro}[theo]{Property}

\theoremstyle{remark}

\theoremstyle{complement}

\newcommand{\pr}{\mbox{P}}

\newcommand{\Corr}{\mbox{Corr}}
\newcommand{\argmax}{\mbox{argmax}}
\newcommand{\eps}{\varepsilon}

\begin{document}

\selectlanguage{english}

\baselineskip 20pt
\vskip 0.32in

\vskip 0.22in
\centerline{\large\bf Inference for changepoint survival models}

\vskip 0.152in
\centerline{\sc Roxane Duroux and John O'Quigley}
\vskip 0.14in
\centerline{\rm Laboratoire de Statistique Th\'eorique et Appliqu\'ee}
\centerline{\rm Universit\'e Pierre et Marie Curie - Paris VI, Paris, France }
 \vskip 0.1in

\noindent {\bf Summary}
We consider a non-proportional hazards model where the regression coefficient is not constant but piecewise constant. Following \citet{andersen-gill1982}, we know that a knowledge of the changepoint leads to a relatively straightforward estimation of the regression coefficients on either side of the changepoint. Between adjacent changepoints, we place ourselves under the proportional hazards model. We can then maximize the partial likelihood to obtain a consistent estimation of the regression coefficients. Difficulties occur when we want to estimate these changepoints. We obtain a confidence region for the changepoint, under a two-step regression model \citep{anderson-senthilselvan1982}, based on the work of \citet{davies1977}. Then we introduce a new estimation method using the standardized score process \citep{chauvel2014}, under a model with multiple changepoints. In this context, rather than make use of the partial likelihood, we base inference on minimization of quadratic residuals. Simulations and an example are provided.

\par
\vskip 0.1in
{\noindent \it Key words}: Cox model, non-proportional hazards, partial likelihood, score process, time-varying effects. \  

\vfill\eject

\section{Introduction}

For the \citet{cox1972} proportional hazards model, the associated hazard can be written
\begin{align}
\lambda(t \vert Z) = \lambda_0(t) \exp(\beta_0^T Z),
\label{mod.PH.Cox}
\end{align}
where $Z \in \R^d$ is a vector covariate, $\lambda_0$ is the unknown baseline hazard, $\beta_0 \in \R^d$ are the regression coefficients for the covariate $Z$, and $a^T$ denotes the transpose vector of the vector $a$. The place of Cox model in the context of regression with censored data is important, especially thanks to its ease of interpretation. We can quote, for instance, the work of \citet{kay1977,kalbfleisch-prentice1980,andersen-gill1982,lin1991}. Some of these papers focus on the asymptotic properties of the partial likelihood estimator, under model \eqref{mod.PH.Cox}, thus enable us to make inference on the parameter of interest $\beta_0$ without worrying about the form of the survival distribution. Nevertheless, model  \eqref{mod.PH.Cox} is not always realistic. We can think, for example, about cancer mortality studies, where the treatment effect decreases in time because the immune system gets used to it. This case is not taken into account by the Cox model which assumes the effects of the covariates to be constant in time. Several authors discuss the case where $\beta_0$ is now a regression function $\beta_0(.)$: \citet{moreau-oquigley-mesbah1985,oquigley-pessione1989,oquigley-pessione1991,liang-self-liu1990,zucker-karr1990,murphy-sen1991,gray1992,hastie-tibshirani1993,verweij-houwelingen1995,lausen-schumacher1996,marzec-marzec1997} just to name a few.

In this paper, we are interested in a particularly simple extension of the Cox model, which is the case where the function $\beta_0$ is piecewise constant. The discontinuities of the function $\beta_0$ are called ``changepoints''. \citet{anderson-senthilselvan1982} investigated the parameters estimation under a two-step regression model, \textit{i.e.}, the estimation of the regression coefficients and the changepoint, in the case of a unique changepoint. Here we extend their analysis and propose an inferential method for the changepoint. Our starting point here is the work of \citet{davies1977}. Following this, we propose an estimation method for a multiple changepoints model with $K$ changepoints, with $K$ fixed in advance.

We begin by introducing the necessary notation in Section \ref{Ch-notations} and present, more formally, the different models. Section \ref{Ch-inference} focuses on the \citet{anderson-senthilselvan1982} model. We recall their estimation method and establish a confidence region for the changepoint. In Section \ref{Ch-modgen}, we place ourselves under the multiple changepoints model with $K$ changepoints and suggest an estimation method using least squares and the standardized score process \citep{chauvel2014}. Simulations for the results of the two previous sections are provided in Section \ref{Ch-simu}. Finally, we illustrate our estimation procedure in Section \ref{Ch-appli} with an application on breast cancer data provided by the Institut Curie, Paris, France.

\section{Notation}\label{Ch-notations}

For all $i \in \{1, \hdots , n\}$, we denote by $(T_i, C_i, Z_i)$ a sequence of independent and identically distributed random variables with the same distribution as the triplet $(T,C,Z)$, where $T$ is the failure time random variable, with a distribution function $F$, $Z \in \R$ is the covariate vector and $C$ is the censoring random variable, independent of $T$ given $Z$. We assume that there exists a real $\tau >0$ such that $[0, \tau]$ is the support of $T$ and $C$, and that these variables follow the model below.
\begin{align}
\lambda(t \vert Z) = \lambda_0(t) \exp\left\{\beta_0(t) Z \right\},
\label{mod.NPH}
\end{align}
In this paper, we focus on two particular models. The first one was introduced by \citet{anderson-senthilselvan1982}. They assumed that the regression coefficient is piecewise constant with two steps. We refer to this model with the term ``single changepoint model''. We can write it the following way.
\begin{align}
\beta_0(t)= \beta_{01} \indi_{t \leq \gamma_0} + \beta_{02} \indi_{t > \gamma_0}, \ \forall t \in [0, \tau],
\label{mod.reduit}
\end{align}
where $\beta_{01}$ and $\beta_{02}$ are real constants and $\gamma_0$, the changepoint, is a positive constant. We call the second model the ``multiple changepoints model''. It can be written
\begin{align}
\beta_0(t)= \beta_{01} \indi_{t \leq \gamma_{01}} + \beta_{02} \indi_{\gamma_{01} < t \leq \gamma_{02}} \hdots + \beta_{0K} \indi_{t > \gamma_{0(K-1)}}, \ \forall t \in [0, \tau],
\label{mod.general}
\end{align}
where $\beta_{01}, \hdots, \beta_{0K}$ are real constants and $\gamma_{01}, \hdots, \gamma_{0(K-1)}$, the changepoints, are positive constants. The number of changepoints $K$ is fixed and known in advance. Notice that the single changepoint model is a particular case of the multiple changepoints model.

For all $i \in \{1, \hdots, n\}$, we define $X_i= \min(T_i,C_i)$ and $\Delta_i= \indi_{T_i \leq C_i}$ such that $X_i$ is the observed time for the patient $i$ and $\Delta_i$ is the assigned status to this patient: ``died'' ($=1$) or ``censored'' ($=0$). We denote by $\mathcal{D}$ the set of right continuous with left limits functions from $[0,\tau]$ in $\R$, and we have $\beta_0 \in \mathcal{D}$. We define, for all $i \in \{1, \cdots ,n\}$ and all $t \in [0, \tau]$, $Y_i(t)= \indi_{X_i \geq t}$. The process $Y_i(t)$ indicates whether the patient $i$ is still at risk at time $t$ ($=1$), or not ($=0$).

In the light of the results obtained by \citet{oquigley-pessione1991}, one approach to inference of the parameter $\gamma_0$ under the single changepoint model, is the use of results developed by \citet{davies1977}. Under model \eqref{mod.reduit}, the log-partial likelihood can be written under the form $L(\beta_1,\beta_2, \gamma)=L_1(\beta_1,\gamma) + L(\beta_2, \gamma)$, where
\begin{align}
L_1(\beta_1, \gamma)= \frac{1}{n} \sum_{X_i \leq \gamma} \Delta_i \left[ \beta_1 Z_i - \log \left\{ \sum_{j=1}^n Y_j(X_i) \exp ( \beta_1 Z_j) \right\} \right] \label{eq-vrais1} \\ 
L_2(\beta_2, \gamma)= \frac{1}{n} \sum_{X_i > \gamma} \Delta_i \left[ \beta_2 Z_i - \log \left\{ \sum_{j=1}^n Y_j(X_i) \exp ( \beta_2 Z_j) \right\} \right]. \label{eq-vrais2}
\end{align}
So, following the steps of \citet{davies1977}, we introduce the statistic $S$ from $\R^2 \times [0,\tau]$ to $\R^2$ defined by
\begin{align*}
S(\beta_1, \beta_2, \gamma) &=  \left( S_1(\beta_1, \gamma) , S_2(\beta_2, \gamma) \right),
\end{align*}
with
\begin{align*}
S_1( \gamma) =  \sqrt{n} V_1(\gamma)^ {1/2} \hat{\beta}_1^ {(n)}(\gamma), \ \
S_2( \gamma) = \sqrt{n} V_2(\gamma)^ {1/2} \hat{\beta}_2^ {(n)}(\gamma),
\end{align*}
where, for $i \in \{1,2\}$, $\hat{\beta}_i^{(n)}(\gamma)$ is the value which maximizes the partial likelihood $L_i(\beta_i,\gamma)$ with $\gamma \in [0, \tau]$ fixed, and
\begin{align*}
V_1(\gamma) = - \frac{\partial^2 L_1}{\partial \beta_1^2} (0,\gamma) = \frac{1}{n} \sum_{i=1}^n \Delta_i \indi_{X_i \leq \gamma} V(0,X_i), \
V_2(\gamma) = - \frac{\partial^2 L_2}{\partial \beta_2^2} (0,\gamma) = \frac{1}{n} \sum_{i=1}^n \Delta_i \indi_{X_i > \gamma} V(0,X_i),
\end{align*}
with
\begin{align*}
V(\beta(t),t) = \frac{S^{(2)}(\beta(t),t)}{S^{(0)}(\beta(t),t)} - \left\{ \frac{S^{(1)}(\beta(t),t)}{S^{(0)}(\beta(t),t)} \right\}^2,
\end{align*}
where, for $r \in \{0,1,2\}$ and all $\beta \in \mathcal{D}$, the functions $S^{(r)}(\beta(t),t)$ are defined as follow
\begin{align}
S^{(r)}(\beta(t),t) = \frac{1}{n} \sum_{j=1}^n Y_j(t) \exp \left( \beta(t) Z_j \right)Z_j^r.
\label{eq-fct.S}
\end{align}
In the following section, inference for the changepoint $\gamma_0$ is based on some results of \citet{davies1977}. Section \ref{Ch-modgen} extends these results to multiple changepoints model \eqref{mod.general}, using the standardized score process \citep{chauvel2014}.

\section{Inference under the single changepoint model}\label{Ch-inference}

For $i \in \{1,2\}$, we denote by
\begin{align*}
\Corr \left\{ S_i(\gamma_1), S_i(\gamma_2) \right\} = \rho_i (\gamma_1 , \gamma_2),
\end{align*}
the correlation coefficient between $S_i(\gamma_1)$ and $S_i(\gamma_2)$. On the segment $[0,\gamma_0)$ on one hand, and on $[\gamma_0, \tau]$ on the other hand, we place ourselves in the case of proportional hazards models. That is why, according to the results of \citet{andersen-gill1982}, we have a convergence in distribution, for $i \in \{1,2\}$:
\begin{align*}
\frac{1}{\sqrt{n}}S_i(\beta_i,\gamma_0) \underset{n \to \infty}{\overset{d}{\longrightarrow}} \mathcal{N}_i(\gamma_0),
\end{align*}
where $\mathcal{N}_i(.)$ is a Gaussian process of mean
\begin{align*}
E_{\beta_{0i}, \gamma_0} \left\{ \mathcal{N}_i(\gamma_1) \right\} = \beta_{0i} V_i(\gamma_0) \rho_i (\gamma_1 , \gamma_0),
\end{align*}
and correlation function
\begin{align*}
\Corr \left\{ \mathcal{N}_i(\gamma_1), \mathcal{N}_i(\gamma_2) \right\} = \rho_i(\gamma_1, \gamma_2),
\end{align*}
for $\gamma_1, \gamma_2 \in [0, \tau]$. We consider now that $n$ is sufficiently large for the deviations of $S_i(\gamma)$ from the Gaussian variable $\mathcal{N}_i(\gamma)$ to be ignored. We also assume that the functions $\rho_i(\gamma_1,\gamma_2)$ are $\mathcal{C}^2$-differentiable.

\subsection{Changepoint estimation}

\citet{anderson-senthilselvan1982} proposed an estimation method, under the single changepoint model \eqref{mod.reduit}, of the regression coefficients $\beta_{01}$, $\beta_{02}$ and the changepoint $\gamma_0$. We recall that the partial log-likelihood is written $L(\beta_1,\beta_2,\gamma) = L_1(\beta_1, \gamma) + L_2(\beta_2,\gamma)$, where the functions $L_1$ and $L_2$ are defined in \eqref{eq-vrais1} and \eqref{eq-vrais2}. A straightforward maximization of $L(\beta_1,\beta_2,\gamma)$ is complex, because the convex optimization methods often need regularity conditions on the function $L$. For example, for the Newton-Raphson method, the function $L$ needs to be $\mathcal{C}^2$-differentiable on $\R^2 \times [0,\tau]$, and this is not the case because of a discontinuity at $\gamma$.

However, we can estimate $\beta_{01}$ and $\beta_{02}$ for every possible value of the changepoint $\gamma$, assuming for instance that it can only occur on a failure time. Then, for $\gamma$ fixed, by maximizing $L_1(\beta_1,\gamma)$ on one hand, and $L_2(\beta_2,\gamma)$ on the other hand, we obtain two processes $\hat{\beta}_{01}(\gamma)$ and $\hat{\beta}_{02}(\gamma)$ verifying
\begin{align*}
\hat{\beta}_{01}(\gamma) = \argmax_{\beta_1} L_1(\beta_1,\gamma), \
\hat{\beta}_{02}(\gamma) = \argmax_{\beta_2} L_2(\beta_2,\gamma).
\end{align*}
Finally, the chosen triplet $(\hat{\beta}_{01}, \hat{\beta}_{02}, \hat{\gamma}_0)$ is defined by the relation
\begin{align*}
(\hat{\beta}_{01}, \hat{\beta}_{02}, \hat{\gamma}_0) = (\hat{\beta}_{01}(\hat{\gamma}_0), \hat{\beta}_{02}(\hat{\gamma}_0), \hat{\gamma}_0) = \argmax_{\gamma} L(\hat{\beta}_{01}(\gamma),\hat{\beta}_{02}(\gamma),\gamma).
\end{align*}
In other words, among all the triplets $(\hat{\beta}_{01}, \hat{\beta}_{02}, \gamma)$ where $\gamma \in \{X_i \ / \ i \in \{1, \hdots, n\}, \ \Delta_i = 1\}$, we choose the one maximizing the partial log-likelihood $L$. Now that we have established the estimation step, we are interested in a confidence region for the changepoint $\gamma_0$. This is detailed in the next section.

\subsection{A confidence region}

The results of \citet{davies1977} were used in the context of a survival problem by \citet{oquigley-pessione1991}. These latter authors studied the model
\begin{align}
\beta_0(t) = \beta_0 \indi_{t \leq \gamma_0} - \beta_0 \indi_{t > \gamma_0}.
\label{mod.OP}
\end{align}
They looked for a test where the null hypothesis was ``$\beta_0=0$'', against the alternative ``$\beta_0>0$''. They proposed a test based on the statistic
\begin{align*}
M=\sup \{ \vert S(\gamma) \vert \ / \ 0 \leq \gamma \leq \tau \},
\end{align*}
where
\begin{align*}
S(\gamma) = \left\{ \frac{\partial L}{\partial \beta}(\beta,\gamma) \right\}_{\beta=0}  \left\{ -\frac{\partial^2 L}{\partial \beta^2}(\beta,\gamma) \right\}_{\beta=0}^{-1/2},
\end{align*}
with $L(\beta,\gamma)$ the partial log-likelihood under the model \eqref{mod.OP}. We make use of this test in order to build a confident region for $\gamma_0$ under the single changepoint model \eqref{mod.reduit}. We use the following statistics
\begin{align}
M_1=\sup \{ S_1(\gamma)  \ / \ 0 \leq \gamma \leq \tau \}, \
M_2=\sup \{  S_2(\gamma)  \ / \ 0 \leq \gamma \leq \tau \}. \label{def-M}
\end{align}

Let $z \in \R$. We detail here the way in which we obtain a confidence region based on the statistic $M_1$. The case of the statistic $M_2$ is similar. Let us define the real $q_\alpha(z,\gamma_0)$ by
\begin{align}
P_{\beta_{01},\gamma_0} \left( \left. M_1=\sup_\gamma S_1(\gamma) > z + q_\alpha(z,\gamma_0) \right\vert S_1(\gamma_0)=z \right) = \alpha.
\label{def-q_alpha}
\end{align}
Then a $1-\alpha$ confident region for $\gamma_0$ is
\begin{align}
\{ \gamma \ / \ S_1(\gamma) > M_1 - q_\alpha(S_1(\gamma_0),\gamma_0) \}.
\label{IC-nul}
\end{align}
The region \eqref{IC-nul} is not usable immediately. So we re-write the left part of \eqref{def-q_alpha}, in order to center and standardize $S_1(\gamma)$ given $S_1(\gamma_0)=z$. We obtain
\begin{align}
P_{\beta_{01},\gamma_0} \left( \left. \sup_\gamma \left[ \frac{S_1(\gamma) - z \rho_1(\gamma,\gamma_0)}{\left\{1-\rho_1(\gamma,\gamma_0)^2 \right\}^{1/2}} - \frac{z + q_\alpha(z,\gamma_0) - z \rho_1(\gamma,\gamma_0)}{\left\{1-\rho_1(\gamma,\gamma_0)^2 \right\}^{1/2}} \right] >0 \right\vert S_1(\gamma_0)=z \right).
\label{eq-avt-IC}
\end{align}
Notice that, when $\rho_1(\gamma,\gamma_0)$ is close to $1$, the second term in \eqref{eq-avt-IC} tends to infinity. Thus, we are only interested in the values of $\gamma$ for which $\rho_1(\gamma,\gamma_0)$ is close to $0$. So we can make the approximation that the first term in \eqref{eq-avt-IC} is independent of $\gamma$, but with a change of sign at $\gamma_0$. So now, we approximate this term by $\text{sgn}(\gamma - \gamma_0) \mathcal{N}_0$, where $\mathcal{N}_0$ is a zero-mean standardized Gaussian variable. The quantity \eqref{eq-avt-IC} becomes
\begin{align*}
P \left( \mathcal{N}_0^2 > \left\{ z + q_\alpha(z,\gamma_0) \right\}^2 - z^2 \right).
\end{align*}
We choose $q_\alpha(z,\gamma_0)$ such that $\{z + q_\alpha(z,\gamma_0)\}^2 - z^2 = \chi_{1,\alpha}^2$, where $\chi_{1,\alpha}^2$ is the $\alpha$ quantile of a chi-squared distribution with one degree of freedom. The approximate $1-\alpha$ confident region for the changepoint$\gamma_0$ is then
\begin{align}
\left\{ \gamma \ / \ S_1(\gamma)^2 > M_1^2 - \chi_{1,\alpha}^2 \right\}.
\label{IC-cool}
\end{align}

\section{Study of the multiple changepoints model}\label{Ch-modgen}

\subsection{Standardized score process}

In this section, we focus on the multiple changepoints model \eqref{mod.general}. Before going any further in the estimation of the $(K-1)$ changepoints $\gamma_{0i}$ and the $K$ regression constants $\beta_{0i}$, we recall some useful notations and results on the standardized score process.

We denote by $N_i(t)= \indi_{T_i \leq t, T_i \leq C_i}$ the counting process and $\bar{N}(t)= \sum_{i=1}^n N_i(t)$. Let $t \in [0, \tau]$, we define the mean and the variance of the covariates $Z$ with respect to the family of probabilities $\{ \pi_i(\beta(t),t)\}_{i \in \{1, \hdots, n\}}$, where
\begin{align*}
\pi_i(\beta(t),t)=\frac{Y_i(t) \exp \left\{\beta(t)Z_i(t) \right\}}{\sum_{j=1}^n Y_j(t) \exp \left\{\beta(t)Z_j(t) \right\}},
\end{align*}
by
\begin{align*}
\mathcal{E}_{\beta(t)}(Z \vert t) = \sum_{i=1}^n Z_i(t) \pi_i(\beta(t),t), \ \mathcal{V}_{\beta(t)}(Z \vert t) = \sum_{i=1}^n Z_i^2(t) \pi_i(\beta(t),t) - \mathcal{E}_{\beta(t)}(Z \vert t)^2.
\end{align*}
The score process $U(\beta,t)$ at time $t \in [0, \tau]$ for the regression function $\beta$ is determined by
\begin{align*}
U(\beta,t) = \int_0^t \left\{ Z_i(s) - \mathcal{E}_{\beta(s)}(Z \vert s) \right\} d \bar{N} (s).
\end{align*}
Notice that, at each failure time, the process $U$ increases by the difference between the covariate value of the dying subject and its mean under the model. At the last failure time, the process equals to the derivative of the partial log-likelihood. \citet{wei1984} proved the convergence of this process to a Brownian bridge, when $\beta$ is the maximum partial likelihood estimate. \citet{haara1987} extended this result to more general cases with non binary covariates.

Next, we consider the standardized score process proposed by \citet{chauvel2014} with some slight modifications. For this, we begin with a time scale change. Let us note $\hat{k}_n= \sharp \{i \ / \ i \in \{1, \hdots, n \}, \ \Delta_i=1 \}$, where $\sharp A$ is the cardinal number of the set $A$, \textit{i.e.}, $\hat{k}_n$ is the number of failure times in the study. According to the strong law of large numbers, we have the almost sure convergence
\begin{align*}
\frac{\hat{k}_n}{n} \overset{p.s.}{\longrightarrow} \alpha_0 = E[\Delta_1] =  \pr (T \leq C).
\end{align*}
We assume that $\alpha_0>0$. This is reasonable because a study never contains only censored data. Furthermore, the law of the iterated logarithm provides us a rate of convergence for $\hat{k}_n/n$ to $\alpha_0$. Indeed, for all $\eps'>0$, for $n$ large enough, almost surely,
\begin{align}
-(1+\eps') \frac{\sqrt{2 \alpha_0 (1-\alpha_0) \log \log n}}{\sqrt{n}} \leq \frac{\hat{k}_n}{n} - \alpha_0 \leq (1+\eps') \frac{\sqrt{2 \alpha_0 (1-\alpha_0) \log \log n}}{\sqrt{n}}.
\label{eq-log-itere}
\end{align}
According to the inequalities \eqref{eq-log-itere}, we can, for all $\eps_0>0$, find an integer $N$ such that, for all $n \geq N$,
\begin{align}
 \left\vert \frac{\hat{k}_n}{n} - \alpha_0 \right\vert \leq \eps_0.
 \label{eq-borne-hatk}
\end{align}
The quantity $\hat{k}_n$ is, by definition, a random variable. In order to manage a theoretical analysis without losing any practical performance, we work with a deterministic sequence $(k_n)_{n \in \N^*}$ having a behaviour close to the one of $(\hat{k}_n)_{n \in \N^*}$, but easier to study. Let $0<\eps<\alpha_0$ fixed. For all $n \in \N^*$, we choose $k_n= \lfloor n(\alpha_0 - \eps) \rfloor \in \N^*$, where $\lfloor x \rfloor$ stands for the integer part of the real $x$. When $n$ goes to infinity, we have the almost sure convergence $k_n/n \rightarrow \alpha_0 - \eps$, and so, for all $\eps_1$, for $n$ large enough, $\vert k_n/n - (\alpha_0- \eps) \vert \leq \eps_1$. We choose $\eps_0=\eps/2$ and $\eps_1=\eps/2$. Then, according to \eqref{eq-borne-hatk}, we can find an integer $N$ such that for all $n \geq N$,
\begin{align*}
\frac{\hat{k}_n}{n} \in [\alpha_0- \frac{\eps}{2}, \alpha_0 + \frac{\eps}{2}] \mbox{ a.s., and } \frac{k_n}{n} \in [\alpha_0- \frac{3\eps}{2}, \alpha_0 - \frac{\eps}{2}].
\end{align*}
Thus, for all $n \geq N$, $k_n \leq \hat{k}_n$ almost surely, $k_n$ takes only non-negative integer values, like $\hat{k}_n$, and behaves approximately like it asymptotically. We can now change the time scale, as proposed by \citet{chauvel2014}:
\begin{align}
\phi_n(X_i)= \frac{\bar{N}(X_i)}{k_n} \left[ 1+ (1-\Delta_i) \frac{\sharp \{j \ / \ j \in \{1, \hdots, n \}, \ X_j<X_i, \ \bar{N}(X_j)=\bar{N}(X_i)\}}{\sharp \{j \ / \ j \in \{1, \hdots, n \}, \ \bar{N}(X_j)=\bar{N}(X_i)\}} \right].
\label{eq-echelle}
\end{align}
After this change, the values $\{0, 1/k_n, 2/k_n \hdots,1\}$ match the failure times and the censoring times are uniformly distributed between the failure times, with respect to their original order. For instance, if $T_1<C_2<C_3<T_4$, then $\phi_n(T_1)<\phi_n(C_2)<\phi_n(C_3)<\phi_n(T_4)$, and $\phi_n(C_2)$ and $\phi_n(C_3)$ are uniformly distributed  between $\phi_n(T_1)$ and $\phi_n(T_4)$. We can define all the useful notions in this scale. We specify the notions in the new scale with a star, \textit{i.e.}, a quantity $x^*$ denotes the quantity $x$ in the new scale \eqref{eq-echelle}. Then, for all $t \in [0,1]$ and $i \in \{1, \hdots,n\}$, $Y_i^*(t)=\indi_{\phi_n(X_i) \leq t}$, $N_i^*(t)= \indi_{\phi_n(X_i) \leq t, \Delta_i=1}$ and
\begin{align*}
\bar{N}^*(t) = \sum_{i=1}^n \indi_{\phi_n(X_i) \leq t, \Delta_i=1}.
\end{align*}
We have now all the tools to define the standardized score process.
\begin{defi}[Standardized score process]
The standardized score process $U^*(\beta(t),t)$ at time $t \in \{0,1/k_n,2/k_n, \hdots,1\}$ for the regression function $\beta$ is defined by
\begin{align}
U^*(\beta(t),t) = \frac{1}{\sqrt{k_n}} \int_0^t \mathcal{V}_{\beta(s)}(Z \vert s)^{-1/2} \{ \mathcal{Z}(s) - \mathcal{E}_{\beta(s)}(Z \vert s)\} d \bar{N}^*(s),
\label{def-proc.score}
\end{align}
where $\mathcal{Z}(.)$ is a left-continuous step function with discontinuities at the points $X_i$ where it takes the value $Z_i(X_i)$. This process is then defined on the whole segment $[0,1]$ by linear interpolation.
\end{defi}
The difference between this definition and the one of \citet{chauvel2014} is the use of $k_n$ instead of $\hat{k}_n$. The necessity of a deterministic sequence $(k_n)_{n \in \N^*}$ for the following property is well-explained by \citet{chauvelphd2014}. We base our changepoints estimation method on this property. The useful assumptions are detailed just below.
\begin{pro}
For all $t \in [0,1]$, under the model \eqref{mod.NPH} and the assumptions {\bf H1-5}, there exist positive constants $C_1(\beta_0)$ and $C_2$ such that the following convergence in probability holds
\begin{align*}
U^*(0,t)- \sqrt{k_n} C_2 \int_0^t \beta_0(s) ds \underset{n \to \infty}{\overset{P}{\rightarrow}} C_1(\beta_0)W,
\end{align*}
where $W$ stands for the standard Brownian motion.
\label{conv-proc.score}
\end{pro}

We denote by $D([0,1],\R)$ the space of right continuous functions with a left limit at every point and endow it with the topology of uniform convergence. We now define, for all $r \in \{0,1,2\}$, the equivalents of $S^{(r)}(\beta(t),t)$, $t \in [0,\tau]$, introduced in \eqref{eq-fct.S}, in the new scale. For all $t \in [0,1]$,
\begin{align*}
S^{(r)}(\beta(t),t) = \frac{1}{n} \sum_{i=1}^n Y_i^*(t) Z_i \left( \phi_n^{-1}(t) \right)^r \exp \left( \beta(t) Z_i \left( \phi_n^{-1}(t) \right) \right).
\end{align*}
Now, we can precise the assumptions of \citet{chauvelphd2014}, sufficient to prove Property \ref{conv-proc.score}. We recall that we place ourselves under the model \eqref{mod.NPH}.
\begin{enumerate}[label= {\bf (H\arabic*)}]
\item (Asymptotic stability) There exist $\delta_1>0$, a neighbourhood of $\beta_0$ of radius $\delta_1$ including the null function, denoted by $\mathbb{B}= \{ \beta, \ \sup_{t \in [0,1]} \vert \beta(t) - \beta_0(t) \vert < \delta_1 \}$, and functions $s^{(r)}$ defined on $\mathbb{B} \times [0,1]$, for $r \in \{0,1,2\}$, such that
\begin{align*}
\sqrt{n} \sup_{t \in [0,1], \ \beta \in \mathbb{B}} \left\vert S^{(r)}(\beta(t),t) - s^{(r)}(\beta(t),t) \right\vert \underset{n \to \infty}{\overset{P}{\longrightarrow}} 0.
\end{align*}
\item (Asymptotic regularity) The deterministic functions $s^{(r)}$, defined in {\bf H1} are uniformly continuous in $t \in [0,1]$ and bounded on $\mathbb{B} \times [0,1]$. Furthermore, for $r \in \{0,1,2\}$, and $t \in [0,1]$, $s^{(r)}(.,t)$ is continuous on $\mathbb{B}$. The function $s^{(0)}$ is bounded away from zero.

We define, for all $t \in [0,1]$ and all $\beta \in \mathbb{B}$, the following quantities.
\begin{align*}
e (\beta(t),t) = \frac{s^{(1)}(\beta(t),t)}{s^{(0)}(\beta(t),t)},
\end{align*}
and
\begin{align*}
v (\beta(t),t) = \frac{s^{(2)}(\beta(t),t)}{s^{(0)}(\beta(t),t)} - e(\beta(t),t)^2.
\end{align*}
\item (Homoscedasticity) For all $t \in [0,1]$ and $\beta \in \mathbb{B}$, $\frac{\partial}{\partial t} v (\beta(t),t)=0$.
\item (Uniformly bounded covariates) There exists $L \in \R_+^*$ such that
\begin{align*}
\sup_{i \in \{1, \hdots,n \}} \sup_{t \in [0,\tau]} \vert Z_i(t) \vert \leq L.
\end{align*}
\item (Non-degenerate variance) There exists a constant $C_\mathcal{V}$ such that, for all $i \in \{1, \hdots, n \}$ verifying $\Delta_i=1$, $\mathcal{V}_0(Z \vert X_i) > C_\mathcal{V}$.
\end{enumerate}
The assumptions {\bf H1-2} are introduced by \citet{andersen-gill1982}. Assumption {\bf H3} is often encountered, in an implicit way, when using proportional hazards models, for the estimation of the variance of the parameter $\beta_0$ or the expression of the log-rank statistic for example. The proof of Property \ref{conv-proc.score} needs {\bf H5} to hold for the $k_n$ first failure times. This is indeed the case when {\bf H5} holds because, by definition of $k_n$, $k_n \leq \hat{k}_n$ almost surely, for $n$ large enough.

\subsection{Changepoints detection}

\begin{figure}[!h]
\centering
\includegraphics[width=\textwidth]{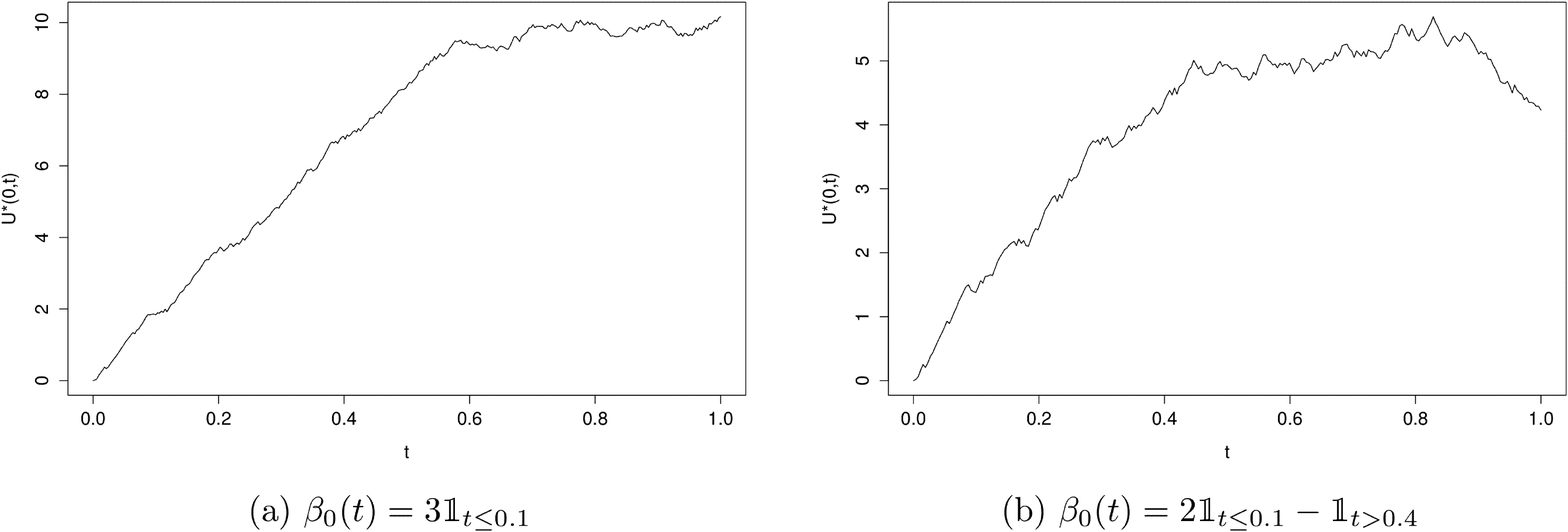}
\caption{Standardized score process}
\label{Ch-fig1}
\end{figure}

Let us start with some illustrations of the process \eqref{def-proc.score} in order to have a better understanding of Property \ref{conv-proc.score}, and to enlighten its interest for the changepoints detection. Figure \ref{Ch-fig1} shows the standardized score process in two cases, both of them are particular cases of the multiple changepoints model \eqref{mod.general}. In both situations, $Z$ follows a uniform distribution in $[0,1]$, $C$ a uniform distribution on $[0,t_c]$ where $t_c$ is set to fix the percentage of censoring approximately at $30\%$, for a data set of $n=500$ observations, and $T$ follows the model \eqref{mod.NPH} for $\lambda_0(t)=1$, for all $t \in [0,\tau]$. For the first situation, we consider a regression function $\beta_0$ such that $\beta_0(t)=3 \indi_{t \leq 0.1}$ for all $t \in [0, \tau]$. For the second one, $\beta_0(t)=2 \indi_{t \leq 0.1} - \indi_{t >0.4} $ for all $t \in [0,\tau]$.

Notice that, in Figure \ref{Ch-fig1}, the deviation expected by Property \ref{conv-proc.score}, which is the integral of the regression function $\beta_0$. We can now see the interest of this process for the changepoints detection. Indeed, if the regression function follows the multiple changepoints model \eqref{mod.general}, then the standardized score process \eqref{def-proc.score} evaluated at the function $\beta(t)=0$ for all $t \in [0, \tau]$ is piecewise affine, according to Property \ref{conv-proc.score}. Thus, in order to find the changepoints $\gamma_{0i}$, we can just carry out a piecewise linear regression. We then obtain the regression constants $\beta_{0i}$ the usual way with the partial likelihood estimates on each step. We recall that the consistency of these estimators to the true regression coefficients, under a proportional hazards model was proved by \citet{andersen-gill1982}.

\subsection{Piecewise linear model}

We provide here some references on changepoints estimation in case of piecewise linear models. We begin with the description of the classical linear regression model.
\begin{align}
y_i = x_i^T \beta + u_i.
\label{reg.lin.classique}
\end{align}
In numerous applications, like the one we are interested in in this paper, it is reasonable to assume that there are $m$ changepoints and so $(m+1)$ segments on which the regression coefficients are constant. In this case,  we can re-write the model \eqref{reg.lin.classique} the following way:
\begin{align}
y_i=x_i^T \beta_j + u_i, \ i \in \{ i_{j-1}+1, \hdots, i_j\}, \ j \in \{1, \hdots, m+1\},
\label{reg.lin.mor}
\end{align}
where $j$ is the number of the segment. \citet{bai1994} gave the basis of changepoint estimation in time series. It was extended to other kinds of changepoints by \citet{bai1997,liu1997,hawkins2001,sullivan2002} and \cite{bai2003} for example. The R package \texttt{strucchange} was proposed by \citet{zeileis2001}. The ideas behind the algorithm, \texttt{breakpoints}, for the estimation of these changepoints are detailed by \citet{zeileis2003} and are based on the minimization of the sum of the squared residuals for the model \eqref{reg.lin.mor}. This is the package we use for our simulations presented in Section \ref{Ch-simu}.

We notice that, in the cited papers, the authors are interested in changepoint estimation, but also in the estimation of the regression coefficients. The least squares estimation is able to handle the changepoint estimation, but we need to go back to the partial likelihood to estimate the regression coefficients $\beta_{0i}$ ourselves under the multiple changepoints model \eqref{mod.general}.

\section{Simulations}\label{Ch-simu}
\subsection{Single changepoint model}

We begin with the study of the confidence region \eqref{IC-cool} behaviour with respect to the sample size $n$, the distributions of $C$ and $Z$, and the model on the regression function $\beta_0$. As suggested in Section \ref{Ch-inference}, we can choose to use the statistic $M_1$ or $M_2$ \eqref{def-M} to determine a confidence region for $\gamma_0$. However, in practice, the higher regression coefficient seems to provide better results. So, we choose to use $M_1$ rather than $M_2$ if $\hat{\beta}_{01} > \hat{\beta}_{02}$, and $M_2$ rather than $M_1$ if $\hat{\beta}_{02} > \hat{\beta}_{01}$. The sample size is fixed at $500$ and $1000$. The distribution of $C$ is exponential with parameter $\mu$, where $\mu$ is chosen to fix the percentage of censoring to $30\%$, $50\%$ or $70\%$. The covariate $Z \in \R$ follows a Bernoulli distribution $\mathcal{B}er$ of parameter $1/2$, a uniform distribution $\mathcal{U}$ on $[0,1]$, a Gaussian distribution $\mathcal{N}$ with mean $1/2$ and variance $1/4$, or an exponential distribution $\mathcal{E}$ with parameter $1/2$. Note that the results of Section \ref{Ch-inference} are established for variables with a support in the segment $[0,\tau]$. However, this assumption sometimes does not hold. So, some scenarios take into account infinite supports. We consider that the regression function $\beta_0$ follows one of the three following scenarios: $\beta_0(t) = \indi_{t \leq 0.3}$, $\beta_0(t) = \indi_{t \leq 0.5}$ and $\beta_0(t) = \indi_{t \leq 0.7}$. For every scenario, $1000$ samples are generated to evaluate the empirical level of the confident region for the changepoint $\gamma_0$. These confident regions are settled for a $10\%$ level. We can see in Tables \ref{Ch-tab1} and \ref{Ch-tab2} that the test behaviour is better for continuous covariates. We also notice a slight improvement of the empirical level when the censoring decreases.

\begin{table}[!ht]
\centering
\caption{Empirical levels for the confident regions of the changepoint (in \%) for covariates with a finite support}
\begin{tabular}{|c|c|c|c|c|c|c|c|}
\cline{3-8}
\multicolumn{2}{c|}{} & \multicolumn{2}{c|}{$\beta_0(t) = \indi_{t \leq 0.3}$} &  \multicolumn{2}{c|}{$\beta_0(t) = \indi_{t \leq 0.5}$} &  \multicolumn{2}{c|}{$\beta_0(t) = \indi_{t \leq 0.7}$} \\
\hline
$n$ & \% censoring & $Z \sim \mathcal{B}er$ & $Z \sim \mathcal{U}$ & $Z \sim \mathcal{B}er$ & $Z \sim \mathcal{U}$ & $Z \sim \mathcal{B}er$ & $Z \sim \mathcal{U}$\\
\hline
500 & 0 & 10.4 & 13.3 & 9.8 & 13.4 & 11.5 & 12.9 \\
500 & 30& 12.6 & 13.2 & 11.3 & 13.7 & 12.6 & 13.0 \\
500 & 50& 12.9 & 13.6 & 13.1 & 13.8 & 12.7 & 13.5 \\
\hline
1000 & 0 & 10.3 & 1.7 & 9.4 & 7.5 & 10.5 & 2.9 \\
1000 & 30& 11.8 & 1.4 & 9.7 & 7.6 & 11.8 & 2.8 \\
1000 & 50& 13.2 & 1.9 & 9.9 & 8.4 & 13.4 & 3.4 \\
\hline
\end{tabular}
\label{Ch-tab1}
\end{table}

\begin{table}[!ht]
\centering
\caption{Empirical levels for the confident regions of the changepoint (in \%) for covariates with infinite support}
\begin{tabular}{|c|c|c|c|c|c|c|c|}
\cline{3-8}
\multicolumn{2}{c|}{} & \multicolumn{2}{c|}{$\beta_0(t) = \indi_{t \leq 0.3}$} &  \multicolumn{2}{c|}{$\beta_0(t) = \indi_{t \leq 0.5}$} &  \multicolumn{2}{c|}{$\beta_0(t) = \indi_{t \leq 0.7}$} \\
\hline
$n$ & \% censoring & $Z \sim \mathcal{N}$ & $Z \sim \mathcal{E}$ & $Z \sim \mathcal{N}$ & $Z \sim \mathcal{E}$ & $Z \sim \mathcal{N}$ & $Z \sim \mathcal{E}$\\
\hline
00 & 0 & 0.4 & 0.6 & 0.3 & 10.4 & 0.8 & 12.4 \\
500 & 30& 0.5 & 0.4 & 0.7 & 11.8 & 1.4 & 13.3 \\
500 & 50& 0.3 & 0.7 & 0.9 & 12.8 & 1.5 & 13.1 \\
\hline
1000 & 0 & 0.2 & 0.7 & 0.6 & 9.9 & 2.4 & 8.4 \\
1000 & 30& 0.1 & 0.6 & 1.2 & 10.1 & 3.7 & 8.3 \\
1000 & 50& 0.4 & 0.7 & 0.8 & 10.5 & 4.2 & 9.3 \\
\hline
\end{tabular}
\label{Ch-tab2}
\end{table}

\begin{figure}[!h]
\centering
\includegraphics[width=\textwidth]{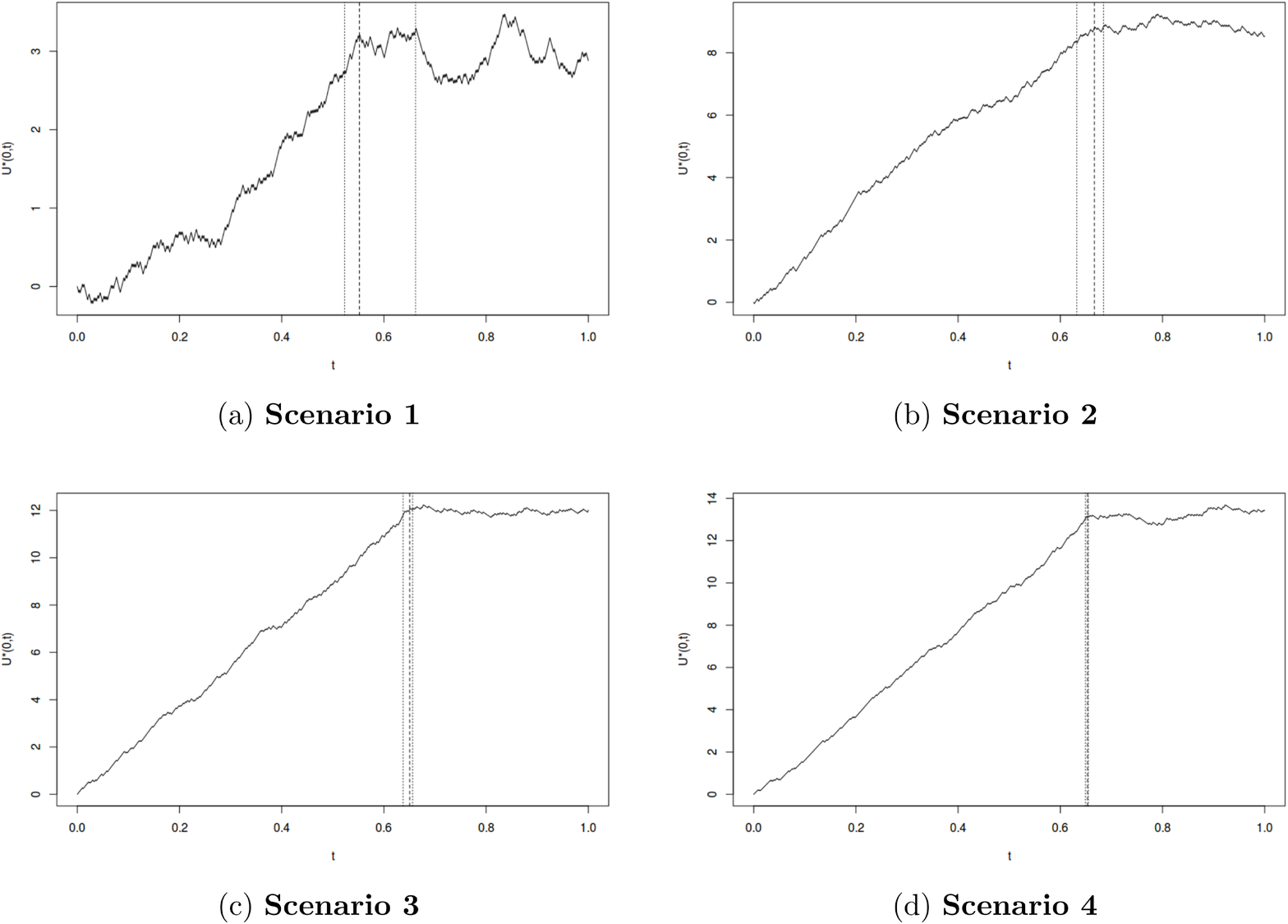}
\caption{Evolution of the $95\%$ confident interval of $\gamma_0$ with respect to the distance $\vert \beta_{01} - \beta_{02} \vert$. Standardized score process (black), estimator value (red), bounds of the $95\%$ confident interval (blue)}
\label{Ch-fig2}
\end{figure}

We can look at the confidence region \eqref{IC-cool} behaviour with respect to the distance between the two regression coefficients, \textit{i.e.}, with respect to $\vert \beta_{01} - \beta_{02} \vert$. An illustration of the obtained results is presented in Figure \ref{Ch-fig2}. In order to obtain these graphics, we generated data sets of size $n=1000$ under the single changepoint model \eqref{mod.reduit} with $\beta_{02}=0$, $\beta_{01}$ fixed at $0.5$, $1$, $1.5$ and $2$, and $\gamma_0$ fixed at $0.5$, $0.4$, $0.3$ and $0.2$. $Z$ follows a uniform distribution on $[0,1]$, $C$ a uniform distribution on $[0,\tau]$, where $\tau$ is fixed to obtain around $30\%$ censoring, and $\lambda_0(t)=1$ for all $t \in [0,\tau]$. We obtain the following estimations:
\begin{itemize}
\item For the model $\beta_0(t) = 0.5 \indi_{t \leq 0.5}$ ({\bf Scenario 1}), we found $\hat{\gamma_0}=0.429$ and $CI_{95\%} = [0.233,0.563]$.
\item For the model $\beta_0(t) = \indi_{t \leq 0.4}$ ({\bf Scenario 2}), we found $\hat{\gamma_0}=0.328$ et $CI_{95\%} = [0.310,0.461]$.
\item For the model $\beta_0(t) = 1.5 \indi_{t \leq 0.3}$ ({\bf Scenario 3}), we found $\hat{\gamma_0}=0.290$ et $CI_{95\%} = [0.275,0.304]$.
\item For the model $\beta_0(t) = 2 \indi_{t \leq 0.2}$ ({\bf Scenario 4}), we found $\hat{\gamma_0}=0.198$ et $CI_{95\%} = [0.196,0.206]$.
\end{itemize}
According to these results, presented in Figure \ref{Ch-fig2}, we can see, that the length of the $95\%$ confidence interval of $\gamma_0$ decreases when the distance between $\beta_{01}$ and $\beta_{02}$ increases. Indeed, it is reasonable to think that, the higher this distance is, the easier it is to find the changepoint of the model. 

\subsection{Multiple changepoints models}

Here, we estimate several changepoints, under the multiple changepoints model \eqref{mod.general}. We start with the study of some single changepoint models \eqref{mod.reduit} and we compare the precisions of the estimation by the method of \citet{anderson-senthilselvan1982}, \textit{i.e.} by maximization of the partial likelihood, and the estimation by the least squares, introduced in Section \ref{Ch-modgen}. Then we study the least squares estimation method in some cases with multiple changepoints. Let us recall that the least squares estimation is carried out with the R package \texttt{strucchange} \citep{zeileis2001}.

For the single changepoint models, we chose an exponential distribution for the censoring $C$, where its parameter is set to fixed the percentage of censoring to $30\%$ or $50\%$. The baseline hazard $\lambda_0$ is identically equal to $1$. The covariate $Z$ follows a uniform distribution on $[0,1]$. Finally, the models on $\beta_0$ are $\beta_0(t) = 0.5 \indi_{t \leq 0.5}$, $\beta_0(t) = \indi_{t \leq 0.4}$ and $\beta_0(t) = 2 \indi_{t \leq 0.3}$. The results are shown in Table \ref{Ch-tab3}. $100$ samples of size $n=1000$ provide us average estimates and empirical standard errors. In Table \ref{Ch-tab3}, ``PL'' stands for ``Partial Likelihood'' and ``LS'' for ``Least Squares''.

\begin{table}[!ht]
\centering
\caption{Comparison of the estimations of the changepoint by partial likelihood and least squares methods. Standard errors in parenthesis.}
\begin{tabular}{|c|c|c|c|}
\hline
Model & \% censoring & PL & LS \\
\hline
$\beta_0(t)= 0.5 \indi_{t \leq 0.5}$ & 0 & 0.648 (0.379) & 0.576 (0.256) \\
$\beta_0(t)= 0.5 \indi_{t \leq 0.5}$ & 30 & 0.689 (0.440) & 0.468 (0.218) \\
$\beta_0(t)= 0.5 \indi_{t \leq 0.5}$ & 50 &  0.670 (0.497) & 0.407 (0.163) \\
\hline
$\beta_0(t) = \indi_{t \leq 0.4}$ & 0 & 0.506 (0.280) & 0.432 (0.136)\\
$\beta_0(t) = \indi_{t \leq 0.4}$ & 30 & 0.438 (0.234) & 0.386 (0.092)\\
$\beta_0(t) = \indi_{t \leq 0.4}$ & 50 & 0.494 (0.341) & 0.329 (0.130)\\
\hline
$\beta_0(t) = 2 \indi_{t \leq 0.3}$ & 0 & 0.307 (0.025) & 0.313 (0.046) \\
$\beta_0(t) = 2 \indi_{t \leq 0.3}$ & 30 & 0.313 (0.043) & 0.304 (0.044) \\
$\beta_0(t) = 2 \indi_{t \leq 0.3}$ & 50 & 0.324 (0.138) & 0.295 (0.091) \\
\hline
\end{tabular}
\label{Ch-tab3}
\end{table}

We notice, with Table \ref{Ch-tab3}, that the least squares method seems to have better performances than the one using the partial likelihood. Indeed, the least squares method loses very little efficiency when the censoring percentage increases and it seems to give a better estimation of the changepoint, after averaging. As expected, for both of the methods, the higher the distance between the two coefficients is, the higher the efficiency is, \textit{i.e.}, the more precise the methods are for the estimation of the changepoint.

\begin{figure}[!h]
\centering
\includegraphics[width=\textwidth]{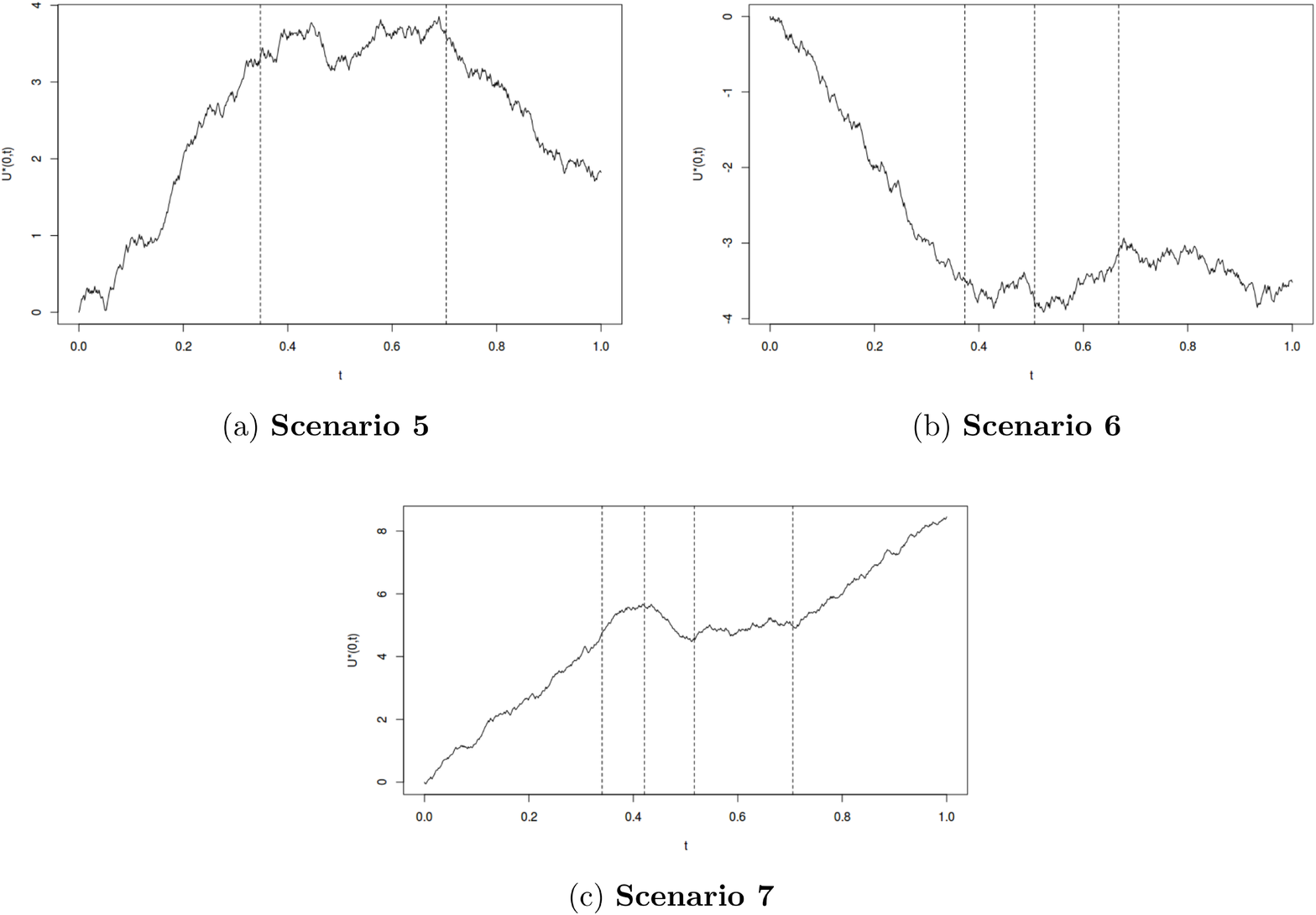}
\caption{Illustration of {\bf Scenarios 5-7} with the standardized score process (black) and the localisation of the changepoints (red)}
\label{Ch-fig3}
\end{figure}

We are now interested in the evolution of the precision of the changepoints estimation by least squares when the number of changepoints increases. We make the censoring rate vary from $0\%$ to $50\%$, as previously. The distribution of $Z$ and the baseline hazard are still the same. The sample size $n$ takes the values $200$, $500$ and $1000$. We generate $100$ samples which enable us to find a mean and a standard error for the changepoints estimates. The studied models are listed below. For a better viewing of these models, we draw, in Figure \ref{Ch-fig3}, the standardized score process for these three models, with $30\%$ of censoring, and a sample size $1000$.
\begin{itemize}
\item {\bf Scenario 5}: $\beta_0(t) = \indi_{t \leq 0.2} - \indi_{t > 0.6} $.
\item {\bf Scenario 6}: $\beta_0(t) = - \indi_{t \leq 0.5} +0.5 \indi_{1.1 < t \leq 2.4} + \indi_{t > 2.4} $.
\item {\bf Scenario 7}: $\beta_0(t) = 2 \indi_{t \leq 0.1} - \indi_{0.2 < t \leq 0.3} + 1.5 \indi_{t > 0.6} $.
\end{itemize}
The results are presented in Table \ref{Ch-tab4} for the {\bf Scenario 5}, Table \ref{Ch-tab5} for the {\bf Scenario 6} and Table \ref{Ch-tab6} for the {\bf Scenario 6}. We can remark once again that the precision in the estimation does not depend on the censoring. We also notice that it increases when the sample size goes from $200$ to $500$. But the difference of precision is negligible when the size goes from $500$ to $1000$ observations. As it can be seen on Figure \ref{Ch-fig3} (b), the last changepoint is close to the end  of the data set. It explains why, in Table \ref{Ch-tab5}, we have a poor estimation of the last changepoint $\gamma_3$.

\begin{table}[!ht]
\centering
\caption{Evolution of the precision of the changepoints estimators in {\bf Scenario 5}. Standard errors in parenthesis.}
\begin{tabular}{|c|c|c|c|}
\cline{3-4}
\multicolumn{2}{c|}{} & \multicolumn{2}{c|}{{\bf Scenario 5}} \\
\hline
$n$ & \% censoring & $\gamma_1$ & $\gamma_2$ \\
\hline
200 & 0 & 0.266 (0.168) & 1.179 (0.633) \\
\hline
200 & 30 & 0.171 (0.094) & 0.694 (0.370) \\
\hline
200 & 50 & 0.128 (0.075) & 0.417 (0.211) \\
\hline
500 & 0 & 0.243 (0.138) & 0.672 (0.255) \\
\hline
500 & 30 & 0.183 (0.104) & 0.665 (0.132) \\
\hline
500 & 50 & 0.135 (0.071) & 0.434 (0.188) \\
\hline
1000 & 0 & 0.218 (0.066) & 0.751 (0.276) \\
\hline
1000 & 30 & 0.200 (0.066) & 0.681 (0.241) \\
\hline
1000 & 50 & 0.143 (0.061) & 0.643 (0.169) \\
\hline
\end{tabular}
\label{Ch-tab4}
\end{table}

\begin{table}[!ht]
\centering
\caption{Evolution of the precision of the changepoints estimators in {\bf Scenario 6}. Standard errors in parenthesis.}
\begin{tabular}{|c|c|c|c|c|}
\cline{3-5}
\multicolumn{2}{c|}{} & \multicolumn{3}{c|}{{\bf Scenario 6}} \\
\hline
$n$ & \% censoring & $\gamma_1$ & $\gamma_2$ & $\gamma_3$ \\
\hline
200 & 0 & 0.407 (0.198) & 0.901 (0.280) & 1.566 (0.429) \\
\hline
200 & 30 & 0.299 (0.160) & 0.703 (0.198) & 1.269 (0.327) \\
\hline
200 & 50 & 0.185 (0.107) & 0.452 (0.166) & 0.857 (0.222) \\
\hline
500 & 0 & 0.428 (0.141) & 0.880 (0.228) & 1.604 (0.433) \\
\hline
500 & 30 & 0.309 (0.144) & 0.679 (0.211) & 1.193 (0.272) \\
\hline
500 & 50 & 0.248 (0.121) & 0.559 (0.155) & 1.026 (0.262) \\
\hline
1000 & 0 & 0.423 (0.138) & 0.864 (0.257) & 1.589 (0.454) \\
\hline
1000 & 30 & 0.305 (0.149) & 0.641 (0.171) & 1.174 (0.268) \\
\hline
1000 & 50 & 0.202 (0.114) & 0.483 (0.135) & 0.875 (0.225) \\
\hline
\end{tabular}
\label{Ch-tab5}
\end{table}

\begin{table}[!ht]
\centering
\caption{Evolution of the precision of the changepoints estimators in {\bf Scenario 7}. Standard errors in parenthesis.}
\begin{tabular}{|c|c|c|c|c|c|}
\cline{3-6}
\multicolumn{2}{c|}{} & \multicolumn{4}{c|}{{\bf Scenario 7}} \\
\hline
$n$ & \% censoring & $\gamma_1$ & $\gamma_2$ & $\gamma_3$ & $\gamma_4$ \\
\hline
200 & 0 & 0.086 (0.065) & 0.307 (0.195) & 0.662 (0.185) & 1.121 (0.264) \\
\hline
200 & 30 & 0.054 (0.029) & 0.168 (0.117) & 0.451 (0.179) & 0.793 (0.183) \\
\hline
200 & 50 & 0.033 (0.017) & 0.093 (0.055) & 0.237 (0.147) & 0.555 (0.172) \\
\hline
500 & 0 & 0.077 (0.040) & 0.291 (0.191) & 0.637 (0.162) & 1.095 (0.304) \\
\hline
500 & 30 & 0.059 (0.033) & 0.175 (0.114) & 0.473 (0.175) & 0.798 (0.176) \\
\hline
500 & 50 & 0.040 (0.024) & 0.103 (0.053) & 0.246 (0.135) & 0.587 (0.134) \\
\hline
1000 & 0 & 0.085 (0.031) & 0.271 (0.157) & 0.577 (0.138) & 1.016 (0.282) \\
\hline
1000 & 30 & 0.059 (0.033) & 0.165 (0.105) & 0.433 (0.170) & 0.755 (0.176) \\
\hline
1000 & 50 & 0.042 (0.022) & 0.101 (0.042) & 0.267 (0.131) & 0.590 (0.111) \\
\hline
\end{tabular}
\label{Ch-tab6}
\end{table}

\section{Application}\label{Ch-appli}

\begin{figure}[!h]
\centering
\includegraphics[width=\textwidth]{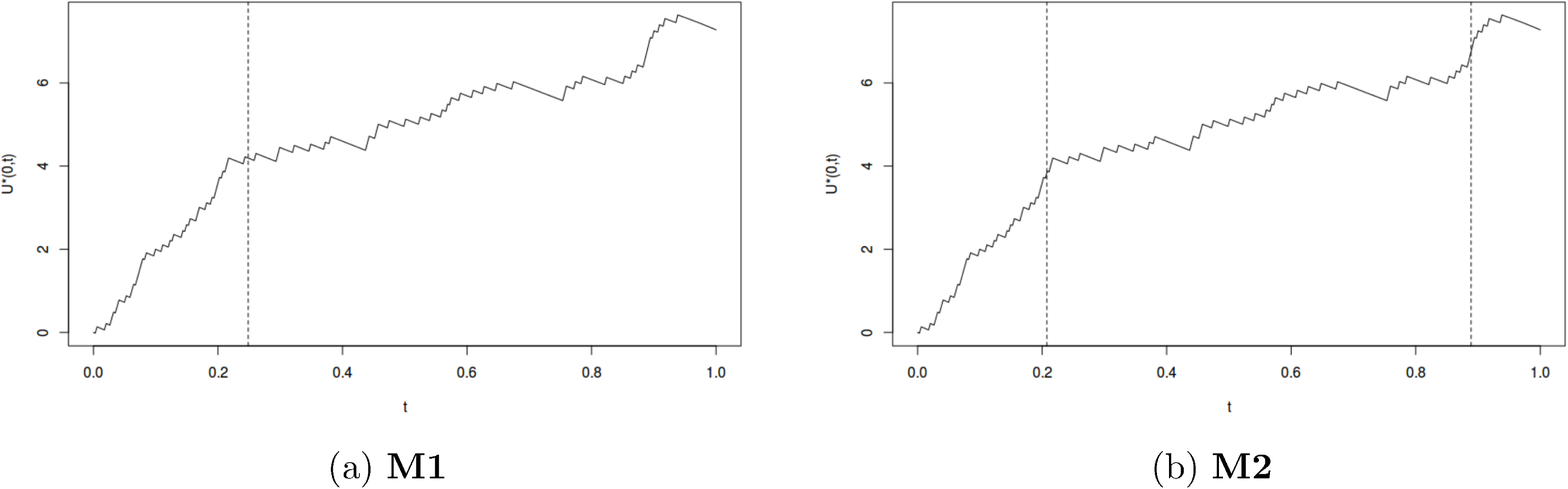}
\label{Ch-appli1}
\caption{Standardized score process for the tumour size covariate (black) and changepoints estimates (red)}
\end{figure}

In this section, we apply the methods of Section \ref{Ch-modgen} and \ref{Ch-simu} on breast cancer data collected at the Institut Curie. In this data set, we have information on the survival time of 1504 patients suffering from breast cancer. The covariates at our disposal are their age, their histological grade, their cancer stage, their tumour size and their progesterone receptor status. We focus here on the covariate ``tumour size''. We split the patients into two groups: the patients with a tumour size smaller than or equal to $60$mm, and the ones with a tumour size higher than $60$mm. We obtain a binary covariate. Figure \ref{Ch-appli1} presents the standardized score process for the binary tumour size covariate. We notice a deviation from linearity and so, we can think of a time-dependent regression coefficient. More precisely, the slope of the process is decreasing, \textit{i.e.}, the tumour size effect diminish through time. According to Figure \ref{Ch-appli1}, we can wonder whether, at the end of the data set, the tumour size effect raises. Thus we hesitate between two models: a single changepoint model \eqref{mod.reduit} with one changepoint, or a model with two changepoints. We use the least squares method for their estimation, then we estimate the regression coefficients on either side of the changepoints by maximization of the partial likelihood. We obtain the following results
\begin{align*}
{\bf M1} & \qquad \hat{\beta}(t) = 1.68 \indi_{t \leq 28.06} + 0.58 \indi_{t > 28.06}\\
{\bf M2} & \qquad \hat{\beta}(t) = 1.82 \indi_{t \leq 26} + 0.64 \indi_{26 < t \leq 73.00} + 1.03 \indi_{t > 73.00}.
\end{align*}
The estimation of these changepoints is also presented in Figure \ref{Ch-appli1}.

\section{Discussion}

In this paper, we are interested in an extension of the Cox model, more precisely, the case where the regression function $\beta_0$ is piecewise constant. We extend the analysis of \citet{anderson-senthilselvan1982} and propose an inference method on the changepoint under a model with a single changepoint. We provide a confidence region for this parameter, based on the work of \citet{davies1977}. We propose an estimation method for a more multiple changepoints model with $K$ changepoints, where $K$ is fixed and known in advance. This method leans upon the standardized score process \citep{chauvel2014}. The simulations show good performances for this latter estimation method, even if the censoring or the number of changepoints is increasing. Two paths may be naturally worthy of further study. At first, it may be useful to develop an inference method based on the standardized score process. Then we would be able to test the null hypothesis ``the model contains $K$ changepoints'' against the alternative ``the model contains $(K+1)$ changepoints''. The second path is obviously to study the changepoints estimation under a model with $K$ changepoints, but where $K$ is unknown, which is a more realistic situation.

\bibliographystyle{natbib}
\bibliography{biblio}

\begin{thebibliography}{30}
\providecommand{\natexlab}[1]{#1}
\providecommand{\url}[1]{\texttt{#1}}
\expandafter\ifx\csname urlstyle\endcsname\relax
  \providecommand{\doi}[1]{doi: #1}\else
  \providecommand{\doi}{doi: \begingroup \urlstyle{rm}\Url}\fi

\bibitem[Andersen and Gill(1982)]{andersen-gill1982}
Andersen, P.~K. and Gill, R.~D.
\newblock (1982).
\newblock Cox's regression model for counting processes: A large sample study.
\newblock \emph{The Annals of Statistics}, 10\penalty0 (4):\penalty0 pp.
  1100--1120.

\bibitem[Anderson and Senthilselvan(1982)]{anderson-senthilselvan1982}
Anderson, J.~A. and Senthilselvan, A.
\newblock (1982).
\newblock A two-step regression model for hazard functions.
\newblock \emph{Applied Statistics}, pages 44--51.

\bibitem[Bai(1994)]{bai1994}
Bai, J.
\newblock (1994).
\newblock Least squares estimation of a shift in linear processes.
\newblock \emph{Journal of Time Series Analysis}, 15\penalty0 (5):\penalty0
  453--472.

\bibitem[Bai(1997)]{bai1997}
Bai, J.
\newblock (1997).
\newblock Estimation of a change point in multiple regression models.
\newblock \emph{Review of Economics and Statistics}, 79\penalty0 (4):\penalty0
  551--563.

\bibitem[Bai and Perron(2003)]{bai2003}
Bai, J. and Perron, P.
\newblock (2003).
\newblock Computation and analysis of multiple structural change models.
\newblock \emph{Journal of applied econometrics}, 18\penalty0 (1):\penalty0
  1--22.

\bibitem[Chauvel(2014)]{chauvelphd2014}
Chauvel, C.
\newblock \emph{Empirical Processes for Inference in the Non-Proportional
  Hazards model}.
\newblock PhD thesis, Université Pierre et Marie Curie - Paris 6, Paris,
  (2014).

\bibitem[Chauvel and O'Quigley(2014)]{chauvel2014}
Chauvel, C. and O'Quigley, J.
\newblock (2014).
\newblock Tests for comparing estimated survival functions.
\newblock \emph{Biometrika}, page asu015.

\bibitem[Cox(1972)]{cox1972}
Cox, D.~R.
\newblock (1972).
\newblock Regression models and life-tables.
\newblock \emph{Journal of the Royal Statistical Society. Series B
  (Methodological)}, 34\penalty0 (2):\penalty0 pp. 187--220.

\bibitem[Davies(1977)]{davies1977}
Davies, R.~B.
\newblock (1977).
\newblock Hypothesis testing when a nuisance parameter is present only under
  the alternative.
\newblock \emph{Biometrika}, 64\penalty0 (2):\penalty0 247--254.

\bibitem[Gray(1992)]{gray1992}
Gray, R.~J.
\newblock (1992).
\newblock Flexible methods for analyzing survival data using splines, with
  applications to breast cancer prognosis.
\newblock \emph{Journal of the American Statistical Association}, 87\penalty0
  (420):\penalty0 942--951.

\bibitem[Haara(1987)]{haara1987}
Haara, P.
\newblock A note on the asymptotic behaviour of the empirical score in cox’s
  regression model for counting processes.
\newblock In \emph{Proceedings of the 1st World Congress of the Bernoulli
  Society}, pages 139--142, (1987).

\bibitem[Hastie and Tibshirani(1993)]{hastie-tibshirani1993}
Hastie, T. and Tibshirani, R.
\newblock (1993).
\newblock Varying-coefficient models.
\newblock \emph{Journal of the Royal Statistical Society. Series B
  (Methodological)}, 55\penalty0 (4):\penalty0 757--796.

\bibitem[Hawkins(2001)]{hawkins2001}
Hawkins, D.~M.
\newblock (2001).
\newblock Fitting multiple change-point models to data.
\newblock \emph{Computational Statistics \& Data Analysis}, 37\penalty0
  (3):\penalty0 323--341.

\bibitem[Kalbfleisch and Prentice(1980)]{kalbfleisch-prentice1980}
Kalbfleisch, J.~D. and Prentice, R.~L.
\newblock \emph{The statistical analysis of failure time data}.
\newblock Wiley series in probability and mathematical statistics: Applied
  probability and statistics. Wiley, (1980).
\newblock ISBN 9780471055198.

\bibitem[Kay(1977)]{kay1977}
Kay, R.
\newblock (1977).
\newblock Proportional hazard regression models and the analysis of censored
  survival data.
\newblock \emph{Applied Statistics}, pages 227--237.

\bibitem[Kleiber et~al.(2002)Kleiber, Hornik, Leisch, and Zeileis]{zeileis2001}
Kleiber, C., K.~Hornik, F.~Leisch, and Zeileis, A.
\newblock (2002).
\newblock strucchange: An r package for testing for structural change in linear
  regression models.
\newblock \emph{Journal of Statistical Software}, 7\penalty0 (2):\penalty0
  1--38.

\bibitem[Lausen and Schumacher(1996)]{lausen-schumacher1996}
Lausen, B. and Schumacher, M.
\newblock (1996).
\newblock Evaluating the effect of optimized cutoff values in the assessment of
  prognostic factors.
\newblock \emph{Comput. Stat. Data Anal.}, 21\penalty0 (3):\penalty0 307--326.

\bibitem[Liang et~al.(1990)Liang, Self, and Liu]{liang-self-liu1990}
Liang, K.~Y., S.~G. Self, and Liu, X.
\newblock (1990).
\newblock The cox proportional hazards model with change point: An
  epidemiologic application.
\newblock \emph{Biometrics}, 46\penalty0 (3):\penalty0 783--793.

\bibitem[Lin(1991)]{lin1991}
Lin, D.~Y.
\newblock (1991).
\newblock Goodness-of-fit analysis for the cox regression model based on a
  class of parameter estimators.
\newblock \emph{Journal of the American Statistical Association}, 86\penalty0
  (415):\penalty0 pp. 725--728.

\bibitem[Liu et~al.(1997)Liu, Wu, and Zidek]{liu1997}
Liu, J., S.~Wu, and Zidek, J.~V.
\newblock (1997).
\newblock On segmented multivariate regression.
\newblock \emph{Statistica Sinica}, 7\penalty0 (2):\penalty0 497--525.

\bibitem[Marzec and Marzec(1997)]{marzec-marzec1997}
Marzec, L. and Marzec, P.
\newblock (1997).
\newblock On fitting cox's regression model with time-dependent coefficients.
\newblock \emph{Biometrika}, 84\penalty0 (4):\penalty0 901--908.

\bibitem[Moreau et~al.(1985)Moreau, O'Quigley, and
  Mesbah]{moreau-oquigley-mesbah1985}
Moreau, T., J.~O'Quigley, and Mesbah, M.
\newblock (1985).
\newblock A global goodness-of-fit statistic for the proportional hazards
  model.
\newblock \emph{Journal of the Royal Statistical Society. Series C (Applied
  Statistics)}, 34\penalty0 (3):\penalty0 212--218.

\bibitem[Murphy and Sen(1991)]{murphy-sen1991}
Murphy, S.~A. and Sen, P.~K.
\newblock (1991).
\newblock Time-dependent coefficients in a cox-type regression model.
\newblock \emph{Stochastic Processes and their Applications}, 39\penalty0
  (1):\penalty0 153 -- 180.

\bibitem[O'Quigley and Pessione(1989)]{oquigley-pessione1989}
O'Quigley, J. and Pessione, F.
\newblock (1989).
\newblock Score tests for homogeneity of regression effect in the proportional
  hazards model.
\newblock \emph{Biometrics}, 45\penalty0 (1):\penalty0 135--144.

\bibitem[O'Quigley and Pessione(1991)]{oquigley-pessione1991}
O'Quigley, J. and Pessione, F.
\newblock (1991).
\newblock The problem of a covariate-time qualitative interaction in a survival
  study.
\newblock \emph{Biometrics}, 47\penalty0 (1):\penalty0 101—115.

\bibitem[Sullivan(2002)]{sullivan2002}
Sullivan, J.~H.
\newblock (2002).
\newblock Estimating the locations of multiple change points in the mean.
\newblock \emph{Computational Statistics}, 17\penalty0 (2):\penalty0 289--296.

\bibitem[Verweij and Houwelingen(1995)]{verweij-houwelingen1995}
Verweij, P. and Houwelingen, H.~Van.
\newblock (1995).
\newblock Time-dependent effects of fixed covariates in cox regression.
\newblock \emph{Biometrics}, 51\penalty0 (4):\penalty0 1550--1556.

\bibitem[Wei(1984)]{wei1984}
Wei, L.~J.
\newblock (1984).
\newblock Testing goodness of fit for proportional hazards model with censored
  observations.
\newblock \emph{Journal of the American Statistical Association}, 79\penalty0
  (387):\penalty0 649--652.

\bibitem[Zeileis et~al.(2003)Zeileis, Kleiber, Kr{\"a}mer, and
  Hornik]{zeileis2003}
Zeileis, A., C.~Kleiber, W.~Kr{\"a}mer, and Hornik, K.
\newblock (2003).
\newblock Testing and dating of structural changes in practice.
\newblock \emph{Computational Statistics \& Data Analysis}, 44\penalty0
  (1):\penalty0 109--123.

\bibitem[Zucker and Karr(1990)]{zucker-karr1990}
Zucker, David~M. and Karr, Alan~F.
\newblock (1990).
\newblock Nonparametric survival analysis with time-dependent covariate
  effects: a penalized partial likelihood approach.
\newblock \emph{Ann. Statist.}, 18\penalty0 (1):\penalty0 329--353.

\end{thebibliography}

\end{document}